# Dynamic interplay between tumour, stroma and immune system can drive or prevent tumour progression

R J Seager[1], Cynthia Hajal[2], Fabian Spill[1,2,*], Roger D Kamm[2,*] and Muhammad H Zaman[1,3,*]

[1]Department of Biomedical Engineering, Boston University, 44 Cummington Mall, Boston MA 02215

[2]Department of Mechanical Engineering, Massachusetts Institute of Technology, 77 Massachusetts Avenue, Cambridge, MA 02139

[3]Howard Hughes Medical Institute, Boston University, Boston, MA 02215

[*]Correspondence may be addressed to FS (fspill@mit.edu), RDK (rdkamm@mit.edu) or MHZ (zaman@bu.edu)

## ABSTRACT

In the tumour microenvironment, cancer cells directly interact with both the immune system and the stroma. It is firmly established that the immune system, historically believed to be a major part of the body's defence against tumour progression, can be reprogrammed by tumour cells to be ineffective, inactivated, or even acquire tumour promoting phenotypes. Likewise, stromal cells and extracellular matrix can also have pro- and anti-tumour properties. However, there is strong evidence that the stroma and immune system also directly interact, therefore creating a tripartite interaction that exists between cancer cells, immune cells and tumour stroma. This interaction contributes to the maintenance of a chronically inflamed tumour microenvironment with pro-tumorigenic immune phenotypes and facilitated metastatic dissemination. A comprehensive understanding of cancer in the context of dynamical interactions of the immune system and the tumour stroma is therefore required to truly understand the progression toward and past malignancy.

## 1. Introduction

A tumour is composed of more than cancer cells alone. It is in fact a diverse microecosystem including immune cells, stromal cells, the extracellular matrix (ECM), all of which have an effect on the progression of the disease. [1].

Cancer cells first interact with their ECM through a combination of mechanical interactions, enzymatic manipulation of the matrix structure, or signalling interactions, leading to changes in the degree of inflammation and hypoxia of the tumour microenvironment as well as its levels of intracellular signalling [2,3]. Cancer cells also directly interact with stromal cells, like fibroblasts, through paracrine signalling and direct cell-cell contact signalling to create favourable physical or molecular signalling environments for continued tumour progression [4].

Second, cancer cells interact with the immune system, provoking both pro-tumorigenic and anti-tumorigenic behaviours [5]. Importantly, the immune system helps maintain a continuous state of inflammation in the tumour microenvironment, promoting cancer cell growth and metastasis as well as angiogenesis in the tumour microenvironment [5].

Finally, cancer-associated fibroblasts (CAFs) and other stromal cells can recruit immune cells to the tumour and influence their behaviour towards cancer cells [6]. Moreover, they are involved in remodelling of the ECM that can lead to further cancer proliferation and metastatic dissemination, as well as influence immune cell behaviour.

In this review, we examine this tripartite interaction to demonstrate that the dynamic interplay between cancer cells, the immune system, and tumour stroma greatly influence the survival, proliferation and metastasis of cancer cells.

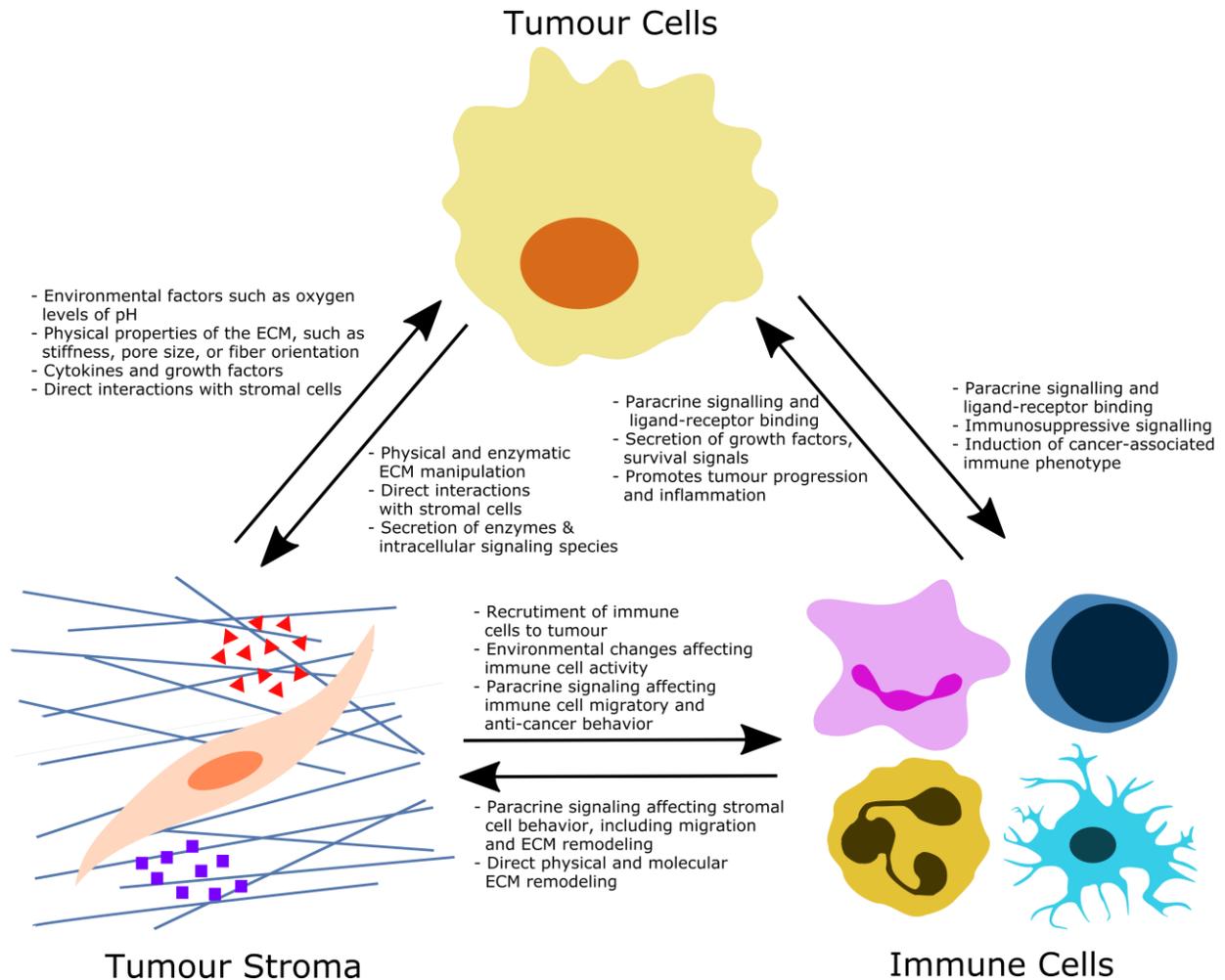

**Figure 1. Tripartite interaction within tumour ecosystem.** Tumour cell behaviour is strongly influenced by the dynamic interactions between the cancer cells themselves, the cells of the immune system, and the tumour stroma—a broad category containing the extracellular matrix (ECM), stromal cells, and intracellular signalling species. The cancer cells, through aberrant signalling and uncontrolled growth, produce a unique microenvironment which results in differential behaviours on the part of both immune and stromal cells. Stromal cells act to reshape the microenvironment in favour of continued tumour progression and express growth factors and inflammatory cytokines. Additionally, cancer cells engage the ECM physically and reshape the ECM mechanically and enzymatically through MMPs and other ECM-remodelling enzymes. Through signalling interactions with immune cells the tumour is able to suppress the anti-cancer immune response and induce the immune cells to secrete pro-inflammatory factors, as well as growth and survival signals which aid tumour progression.

## 2. Cancer Interactions with the Tumour Stroma

Cancer cells constantly interact with their surrounding environment. They alter their behaviour in response to changes in their microenvironment, but also modify their environment to create a more physically, chemically, and metabolically hospitable niche conducive to continued disease progression [7–9]. Directly and indirectly, many of these changes are effected though interactions between cancer and the tumour stroma, here defined as the non-immune, non-neoplastic tumour constituents, in particular the ECM, paracrine signalling molecules such as cytokines, and tissue-supporting cells such as fibroblasts [4].

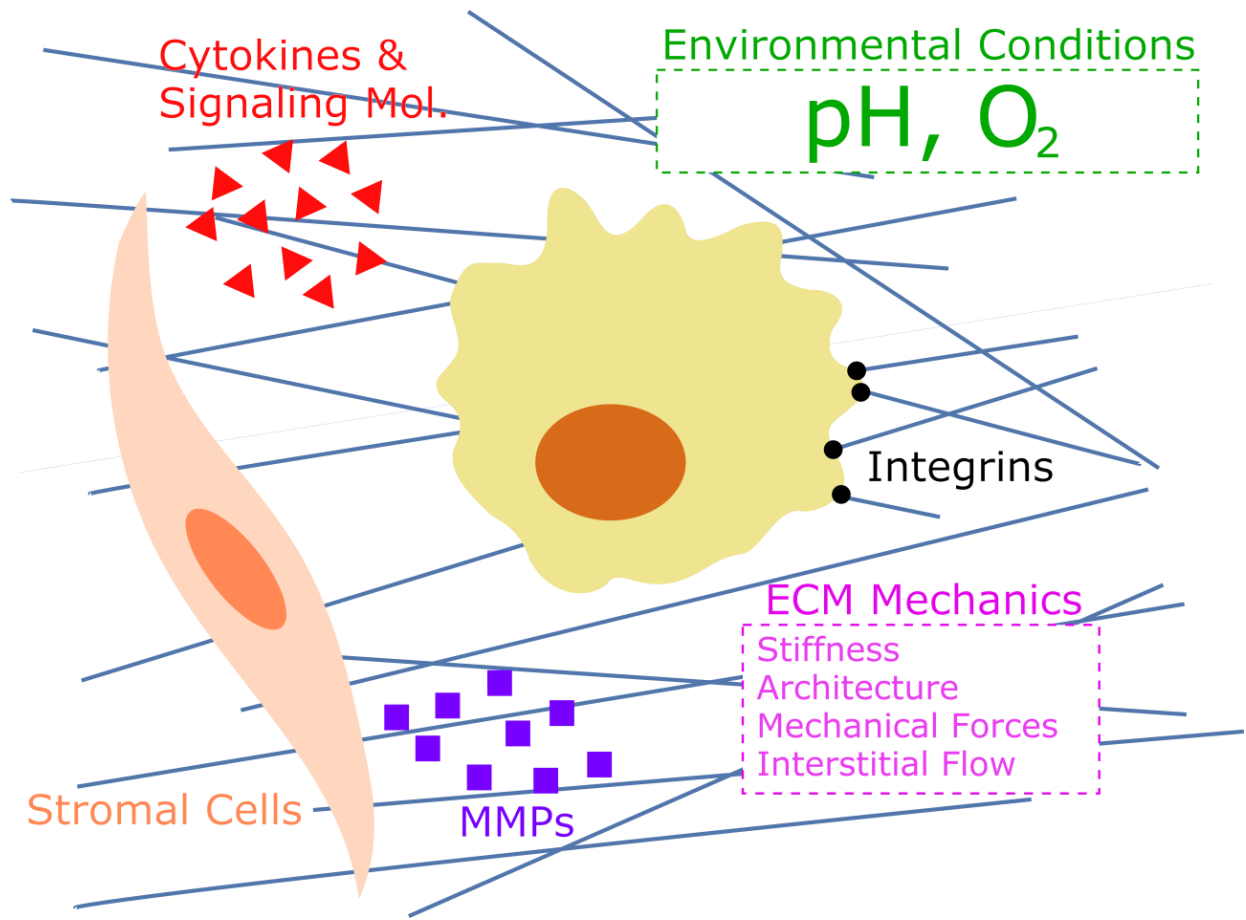

**Figure 2. Summary of stromal conditions, species, and cells affecting and affected by cancer cell activity.** Within the tumour stroma, environmental conditions, mechanical forces and extracellular matrix (ECM) structure, stromal cell activity, and the activity of various extracellular signalling molecules and enzymes all affect and are affected by cancer cell activity. Environmental conditions like pH and microenvironmental oxygen content affect cancer cell metabolism, but are also heavily influenced by cancer cell activity. Cancer cells also mechanically interact with the surrounding ECM and modulate their behaviour according to ECM mechanical properties such as stiffness and fibre architecture, but also respond to dynamic stimuli such as mechanical forces conducted through the ECM and interstitial flows and pressures. Furthermore, the integrins through which the cancer cells form focal adhesions with the ECM structure can initiate signalling pathways within the cancer cells in addition to conduction mechanical forces. The activity of stromal cells, such as the fibroblast pictured above, are heavily affected by cancer paracrine signalling in the tumour microenvironment, but also can change the physical and signalling characteristics of the microenvironment, both of which are significant determinants of cancer cell behaviour. The activity of signalling molecules, such as the pictured cytokines but also growth and survival factors, within the tumour microenvironment are also important determinants of not only cancer cell behaviour but also of stromal and immune cell (not pictured) behaviour. Finally, enzymatic activity, such as that of matrix metalloproteinases (MMPs), can shape the tumour microenvironment both physically and chemically, both of which can greatly affect cancer cell behaviour.

*2.1. Environmental Conditions*

Reduction of oxygen level, known as hypoxia, and decreases of pH, commonly occur in tumours and both have a significant impact tumour progression as well as on interactions of the tumour with stromal and immune cells [10,11]. Rapid growth and poor vasculature development commonly create hypoxic microenvironments in the tumour interior, which lead to metabolic changes in cancer cell behaviour, as well as an increase in malignancy. Notably, cancer cells tend to disproportionately rely on anaerobic mitochondrial energy production—a metabolic shift known as the Warburg effect [12]. The observed increased malignancy often results from the influence of hypoxia-induced signalling pathways, such as hypoxia inducible factor (HIF)-1α-signalling, which is correlated with increased tumour angiogenesis, increased metastatic behaviour, and lowered patient survival rates [13–15]. Finally, hypoxia can also induce ECM modelling through HIF-1's induction of P4HA1, P4HA2, and PLOD2 expression in fibroblasts [16].

Indeed, a hypoxic microenvironment has been shown to induce the HIF-1α and HIF-2α-dependent expression of AlkB homolog 5 (ALKBH5), a demethylase associated with the regulation of mRNAs related to embryonic stem cell pluripotency. In breast cancer, this hypoxia-induced signalling process lead to the demethylation NANOG mRNA and the expression of the pluripotency factor it encodes, resulting in the induction of a cancer stem cell (CSC) phenotype [17].

In response to a hypoxic microenvironment and the subsequent expression of HIF-1, the expression of inflammatory mediators, including toll-like receptors 3 and 4 (TLR3 and TLR4) was increased through direct promoter binding by HIF-1. TLR3 and TLR4, when activated, stimulate the expression of HIF-1 through the NF-kB pathway, thus creating a positive feedback loop that works to maintain this inflammatory microenvironment, favourable for cancer progression [18]. In addition, in certain breast cancers, hypoxia-induced HIF expression contributed to chemotherapy resistance, enriching the breast cancer stem cell population through interleukin (IL)-6 and -8 signalling and increasing the expression of multidrug resistance 1 in response to chemotherapy-induced hypoxia [19].

Hypoxia can promote immune escape as well, through the HIF-1α-induced expression of programmed cell death ligand-1 (PD-L1) (also known as B7-H1). PD-1's interaction with its receptor PD-1 is one of a class of immunosuppressive signalling pathways called immune checkpoints, which help maintain self-tolerance and protect the body's tissues from damage during the normal immune response [20]. When this factor binds with its corresponding receptor (PD-1) on cytotoxic T lymphocytes, it causes the apoptosis of the T cells, thus allowing the cancer cells to avoid the cytotoxic T cell immune response [21].

Finally, in response to tumour-like microenvironments characterized by hypoxia and a lack of available nutrients, epigenetic markers can be adjusted, resulting in different cancer cell behaviours for the same cells. For instance, in a hypoxic tumour environment, the expression of the inflammatory mediator protein ANGPTL2 is stimulated by the microenvironment-induced demethylation of its promoter. This results in the promotion of an invasive phenotype, mediated by aberrant integrin and MMP expression [22].

Intratumoural pH can also be a determinant of cancer cell behaviour, as Otto Warburg demonstrated in 1930 when he showed that cancer cells perform fermentative glycolysis in the absence of oxygen to generate adequate energy to sustain their growth [23]. This process promotes the accumulation of lactic acid in the poorly vascularized tumour microenvironment. Combined with a production of $CO_2$ as the end-product of cellular respiration, substantial acidosis is generated at the tumour site, resulting in considerable cellular stress applied. Not only is this stress overcome by cancer cells, but it is also used to promote their expansion and invasion [24].

*2.2. Physical Interactions with the ECM and Microenvironment*

Homeostatic maintenance of tissue health and function is dependent on a continuous process of interaction and feedback between cells and the ECM, composed of a wide variety of structural and functional proteins [25,26]. In particular, the physical characteristics of the ECM can have a significant effect on cell behaviours such as growth and migration [7,27,28].

*2.2.1. Stiffness.* Tumour cells engage in demonstrably different migratory behaviours depending on the stiffness and stiffness gradients observed in their surrounding microenvironment [29]. Two processes responsible for matrix stiffening are matrix deposition and matrix crosslinking [8,30]. For example, in breast cancer, collagen crosslinking accompanies tumorigenesis along with the aforementioned ECM stiffening and increased focal adhesions. Conversely, inducing the crosslinking of an extant collagen matrix results in a stiffer ECM characterized by increased focal adhesions and a more invasive phenotype for the embedded cancer cells [30]. It has been observed that increased collagen density in the breast cancer ECM significantly promotes tumour formation and results in tumours exhibiting a more invasive phenotype [31]. Mechanically, this increase in collagen density creates a stiffer matrix, which promotes invasive migratory behaviours [32,33]. Examining these molecular processes revealed that in regions of higher matrix density and therefore stiffness, cells exhibit increased focal adhesion formation and heightened activity of the focal adhesion kinase (FAK)-Rho signalling loop. This signalling cascade results in a hyperactivation of the Ras-mitogen-activated protein kinase (MAPK) pathway, which results in increased proliferative behaviour in the cancer cells [32].

*2.2.2. Architecture/Alignment.* The stromal ECM is mainly composed of fibrillary glycoproteins including fibronectin and collagens I and III and is often organized into a random, isotropic fibre network [2]. In cancer, this architecture is often changed through mechanical forces and aberrant CAF activity, resulting in a higher degree of fibre alignment in a direction generally normal to the tumour front. These fibres provide conduits for cancer cell migration away from the tumour, and thus, the orientation of the fibres in the ECM surrounding cancer cells is a determinant of their migratory behaviour. In particular, cancer cells exhibit preferential migration in the direction dictated by the ECM architecture [14]. Furthermore, migration and invasive behaviour by cancer cells can actually be blocked by strategically altering the orientation of the ECM [34].

*2.2.3. Force/Stress/Pressure Sensing.* During cell migration, invasive cancer cells demonstrate a sensitivity to traction stresses caused by their internal contraction. When these stresses increase due to increases in tissue stiffness, the cells alter their migration mode in response, shifting from a bleb-based migration mode relying purely on cellular deformations enabling the cell to squeeze through pores in the ECM to a protease-dependent migration mode in which invadopodia-like membrane protrusions are formed [35,36]. Another example of mechanical forces dictating cancer cell behaviour is applied pressures. In the case of metastatic tumours in bone tissue, this pressure sensing is often mediated by osteocytes, the primary mechanotransducing cells in bone tissue, which increase production of MMPs and CCL5, a chemokine known to attract immune cells, in response to increased stresses [37]. These examples demonstrate that this mechanosensitivity can be vested in the cancer cells themselves or occur as a result of an interaction with non-cancerous cells elsewhere in the tumour microenvironment.

Changes in cell migratory and contractile behaviour can also physically alter the ECM structure. One of the most apparent signs of contraction-induced ECM remodelling is the alignment of ECM fibres in a direction generally normal to the tumour front, a phenomenon which also creates pathways for future metastatic excursions [38]. In addition, both collective and individual cell migration induce strain stiffening in the ECM, densifying and aligning the ECM along their migratory path [39]. Local invasion originating at these tumours is largely oriented along these aligned collagen fibres, which also suggests that the degree of tumour ECM collagen alignment could be an indicator of cancer invasiveness [40]. Indeed, in LKB1 mutant cancer cells, this increased alignment has been shown to correspond to high invasiveness [38]. In lung cancer, this process of tumour-centred matrix alignment has been observed at the signalling level where the disruption of the LKB1-MARK1 pathway results in increased ECM remodelling [41].

Distinct from contraction-induced forces is solid stress. Solid stress is the accumulated mechanical stress exerted on surrounding tissue due to tumour growth, as transmitted by the solid and elastic elements of the cells and ECM [42]. These forces are significant only in tumours, as opposed to normal tissue, and furthermore are not correlated with tissue mechanical properties, such as stiffness [42]. Furthermore, this solid stress can differ between primary tumours and metastases, and is positively correlated with tumour size. Finally, the normal surrounding tissue has also been observed to contribute to this stress [42].

*2.2.4. Interstitial flow.* Returning to the idea of tumour growth-induced forces, during tumour growth the mechanical force exerted on surrounding tissue creates stress gradients within the tumour, and as plasma filtrate leaks from malformed blood vessels resulting from tumour-induced angiogenesis, it also elevates interstitial fluid pressure within the tumour [43,44]. This gradient in fluid pressure promotes the flow of interstitial fluid through the tumour and tumour stroma at higher than normal physiologic rates [44]. This interstitial flow can influence the migratory behaviours of tumour cells, suggesting an influence on tumour growth and metastasis [45]. Additionally, interstitial flow induces CAF migration and related ECM remodelling, which in turn results in the migration of cancer cells, suggesting that in vivo, similar conditions will result in increased metastatic activity [46].

*2.3. Molecular Interactions of Cancer Cells with the ECM*

As cancer cells interact with and respond to their environment, they also actively remodel the surrounding ECM. This typically occurs through a combination of actomyosin-mediated mechanical contraction, protease-facilitated matrix degradation, the synthesis of new matrix, and actin polymerization-supported cell protrusion and matrix deformation [2,8].

*2.3.1. ECM Remodelling.* The dynamics of the enzymatic remodelling process, often mediated by MMPs, depend on both the initial physical properties of the ECM at the time of tumour initiation and the invasiveness of the tumour cells themselves, which can often be characterized in terms of such cells predisposition and general capacity for proteolysis [47]. For example, when MMP-9 and Tenascin-C, a tumour ECM protein which influences cell behaviour by binding with cell surface receptors, are co-expressed in cancer, the clinical prognoses for such cases are significantly lowered, even more so than cases in which either species was individually expressed. This suggests that the ability to bind to and interact with the surrounding ECM, as conferred by membrane proteins like tenascin-C, and the ability to enzymatically remodel the surrounding cell membrane, as conferred by proteases like MMP-9, are both heavily implicated in the invasive potential and overall metastatic behaviours of cancer [48].

Finally, these ECM remodelling behaviours are often the result of metabolic aberrations resulting from mutations in the cancer cells themselves. For instance, in lung cancers, LKB1 loss-of-function mutations impair the regulation of lysyl oxidase (LOX) which in turn causes excess collagen deposition in the tumour ECM. This results in the increased activation of $\beta 1$ integrin signalling which in turn promotes cancer cell proliferation and invasion [49]. Furthermore, LOX-mediated collagen cross-linking provokes fibrosis-enhanced tumour cell proliferation and growth, resulting in increased metastasis [50].

*2.3.2. Signalling Interactions with the ECM.* Understanding how the integrins help convert mechanical stimuli from the surrounding environment into phenotypic changes within the cancer cell is essential to understanding the complex and bidirectional interactions between cancer and the tumour microenvironment [51]. By their transmembrane nature, integrins link intracellular signalling of the cell with external physical interactions as they bind with extracellular matrix. This relationship, however, is not unidirectional, and thus, intracellular signalling and mechanics may affect the extracellular matrix through integrin signalling [51].

In cancer cells and healthy cells alike, integrins may activate intracellular signalling pathways in response to extracellular ligand-binding in a process known as "outside-in" signalling [51]. This allows cells to respond to their external environment through the activation of internal signalling pathways dictating cell behaviour [52]. Specifically, in outside-in signalling, the tyrosine kinase focal adhesion kinase (FAK) is recruited to physically engaged integrins, which enhance FAK's kinase activity and trigger other downstream signalling cascades [53]. These specific signalling cascades and outcomes can vary, but often in cancer biology, changes in migratory behaviours are of special interest due to their effect on metastatic potential. For example, in lung cancer cells, outside-in signalling through integrin $\beta 1$ has been shown to drive cell invasion in response to focal adhesion formation with the surrounding ECM [54]. Furthermore, Gα-interacting, vesicle-associated protein (GIV)—a protein required for outside-in signalling—is upregulated, resulting in a heightened activation of trimeric G proteins in response to integrin binding, which in turn enhances PI3K signalling and tumour cell migration. Furthermore, in cancer, GIV is part of a positive feedback loop which enhances integrin-FAK signalling [55].

Cancer cells, including blood borne circulating tumour cells, have also been observed to engage in "inside-out" activation, whereby intracellular signals regulate the ligand affinity of integrins through interactions with the cytoplasmic and transmembrane domains of the integrin proteins [56]. These cytoplasmic actors which can both affect and be affected by integrin binding activity include p130Cas, Src, and talin, all of which have been shown to regulate carcinoma invasion and chemoresistance in carcinomas [57]. In glioblastoma, integrins (αvβ3, αvβ5, αvβ6, and αvβ8) have been shown to control transforming growth factor (TGF)-β signalling, which is known to be a central mediator of that type of brain cancer [58]. Indeed, human metastatic tumours have been observed to overexpress integrin β1.

In addition, in cancer, integrins are sometimes able to activate intracellular signalling pathways, including those promoting cancer progression, without any bound ligand. For example, in breast, lung, and pancreatic carcinomas, integrin αvβ3 is able to recruit KRAS and RalB to the cancer cell membrane in its unoccupied state, resulting in the activation of TBK1 and NF-kB signalling which promotes tumour initiation, ECM detachment, stemness, and resistance to epidermal growth factor receptor (EGFR) inhibiting drugs [59]. Moreover, in non-small cell lung carcinoma (NSCLC), integrin β1 has be shown to be an essential regulator of EGFR expression and silencing it leads to deregulated EGFR signalling in cancer, including increased proliferation, inhibition of apoptosis, increased cell motility, and invasiveness [60]. Moreover, elevated integrin β1 expression has been linked to increased metastatic behaviours in head and neck squamous cell carcinomas [61].

In addition to their structural remodelling of the membrane collagen fibres, MMPs also play important roles in a large number of signalling pathways affecting cell behaviour and response to external conditions, making them vital for healthy tissue growth, but also cancer progression and metastasis [62,63]. Various anticancer drugs have been shown to owe their anticancer and antimetastatic effects to the downregulation or blocking of MMP activity in cancer cells [64]. MMP-9 in particular is a strong promoter of tumour progression, driving the induction of a metastatic phenotype in certain breast cancers [65]. Furthermore, increased MMP expression (MMP-1,-9,-11,and -13) has been observed in higher grade breast cancer, and differential MMP expression has been shown to be significantly correlated with both histological cancer grade and patient outcome [66]. However, although MMP-mediated ECM degradation is an important component of invasive cancer cell behaviour, excessive MMP expression has in fact been shown to inhibit migratory activity with respect to cells with lower expression levels, suggesting the existence of an optimum expression level of MMPs required to maintain an invasive phenotype [67].

Tumour-associated neutrophils (TANs) have been observed to secrete a unique form of MMP-9 which promotes tumour angiogenesis and cancer cell intravasation. This MMP is released as a proenzyme which must be processed proteolytically and activated in order to become functional. In most cases, the proenzyme for MMP-9 is expressed as a complex with tissue inhibitor of metalloproteinase (TIMP)-1, which inhibits MMP-9 activation. Neutrophils, however, do not express TIMP-1, and as a result, MMP-9 is more easily activated, and more easily available to induce tumour progression and metastasis [68]. In the tumour microenvironment, neutrophils represent the primary source of MMP-9, producing more proMMP-9 than even tumour-associated macrophages (TAMs), the traditionally theorized chief producer [69].

*2.4. Inflammation and Cancer*

*Interplay between tumour, stroma, and immune system can affect tumour progression*

In the tumour microenvironment, chronic inflammation is a common feature of many cancers, and manifests as a self-sustaining process promoting continued signalling and immune suppression [70]. Moreover, inflammation is mediated through the microenvironmental circulation of cytokines, chemokines, and growth factors , which, in the context of cancer, mediate the paracrine signals between the tumour and noncancerous stromal tumour components [71]. When cytokine regulation in the tumour microenvironment is disrupted, the likelihood of tumour progression greatly increases and immune activity is modulated or even suppressed.

*Interplay between tumour, stroma, and immune system can affect tumour progression*

**Table 1.** Examples of cytokine signalling in cancer and its effects on cancer cell behaviour.

| Cytokine | Cancer Type | Observed Effects |
| --- | --- | --- |
| IL-1β | Colorectal Carcinoma | Promotes cancer stemness and invasiveness through the activation of Zeb1, a protein associated with the epithelial-mesenchymal transition (EMT) [72]. |
| IL-17 | Colorectal Carcinoma | Overexpressed by tumour-associated macrophages (TAMs) and Th17 (helper T cells) [73]. Associated with an inflammatory microenvironment characterized by high microvessel density and VEGF-mediated angiogenesis by the cancer cells. IL-17 expression levels were found to be a statistically significant factor correlated with patient survival rates [73]. |
| IL-6 | Colorectal Carcinoma | T helper type 17 (Th17)-associated cytokine. Produced in large amounts by tumour infiltrating leukocytes, inducing STAT3 and NF-kB-induced cancer cell growth [74]. |
|  | Osteosarcoma | Induces increased expression of intercellular adhesion molecule-1 and increases cancer cell migration, as mediated by the integrin-linked kinase (ILK/Akt/AP-1) pathway (IKL and Akt activate AP-1) [75]. This contributes to the metastatic potential of these cells by increasing migration and contributing to anchorage-independence of cell colonies by allowing them to attach to each other rather than the ECM [75]. |
| IL-8 | Carcinoma (breast, lung, pancreatic) | Proinflammatory CXC cytokine implicated in the EMT and, thus, the induction of invasive behaviours in tumours [76]. |
|  | Melanoma | When expressed by tumour cells, promotes tumour cell motility, proliferation, and survival in an autocrine manner, while inducing the angiogenesis of epithelial cells and the recruitment of tumour-associated neutrophils (TANs) to the tumour in a paracrine manner [77]. This is an important example of a tumour-secreted factor influencing both tumour behaviour and the behaviour of the noncancerous tumour stroma [77]. |
| TNFα | Melanoma, Breast Carcinoma | Expressed by TAMs. Shown to be important determinant of the migratory behaviours of cancer cells, and thus their propensity for invasive behaviour [41]. |
|  | Colorectal Carcinoma | T helper type 17 (Th17)-associated cytokine. Produced in large amounts by tumour infiltrating leukocytes, inducing STAT3 and NF-kB-induced cancer cell growth [74]. |
| TGF-β | Melanoma, Breast Carcinoma | Expressed by TAMs. Shown to be important determinant of the migratory behaviours of cancer cells, and thus their propensity for invasive behaviour [41]. |

*Interplay between tumour, stroma, and immune system can affect tumour progression*

*2.5. Cancer Interactions with Stromal Cells: Fibroblasts*

The interactions between cancer and stromal cells have been shown to facilitate biological and anatomical changes in the tumour microenvironment. Cancer-associated stromal cells are not themselves cancerous, but do exist under the influence of nearby tumour cells, and often provide tumour-supporting functions, including maintaining a favourable physical, chemical, and biological tumour microenvironment for continued cancer progression [4,78,79].

Much of the physical remodelling of the surrounding ECM in cancer is facilitated through the action of CAFs [80]. In scirrhous gastric carcinoma (SGC)--a highly invasive and metastatic gastric cancer--stromal fibroblasts (SF) form cellular aggregates when in a process characterized by extensive actomyosin-mediated mechanical ECM remodelling by the SFs. This remodelling of the ECM helps to facilitate the invasion and metastasis of the cancer cells themselves [81]. Furthermore, in response to interstitial flow, CAFs can engage in TGF-β1 and collagenase-dependent migration as well as ECM remodelling through Rho-dependent contractility, the latter of which enhances cancer cell invasion [46].

Yet, the contributions of CAFs to cancer cells do not end with ECM remodelling, as they also express pro-tumour signalling factors to create a favourable environment for cancer progression. For instance, in breast carcinomas, normal stromal fibroblasts acquire, through their interactions with cancer cells, two autocrine signalling loops mediated by the cytokines TGF-β and SDF-1, which facilitate resident fibroblast differentiation into an activated species known as myofibroblasts [82]. Phenotypically in between fibroblasts and smooth muscle cells, myofibroblasts exhibit strong tumour-promoting activities, such as angiogenesis in mouse gastric cancer [82,83]. Under hypoxic conditions, CAFs have also been shown to express the marker CD44, which has been implicated in the maintenance of the stemness and malignancy of these tumour-initiating cancer stem cells [84].

Another of the manifold benefits of CAFs to tumour progression is the maintenance of favourable chemical conditions in the tumour microenvironment. CAFs have been observed to overexpress carbonic anhydrase IX (CAIX), an enzyme which drives the acidification of the tumour microenvironment. It is required to induce the activation of stromal fibroblast-delivered MMPs which themselves induce the epithelial-mesenchymal transition--a key stage in tumour progression which promotes migratory and invasive tendencies [85]. Furthermore, in breast cancer, the secretion of the cytokine TGF-β, whether originating in cancer cells or CAFs, promotes tumour progression. It shifts the CAFs in the tumour stroma toward catabolic metabolism and produces building blocks required for mitochondrial metabolism and anabolic growth of tumour cells [86]. Finally, MMP expression in the stroma has also been shown to be an important determinant of cancer cell behaviour. The expression of MMP2 in CAFs is positively correlated with the expression of genes associated with fibroblast activation and matrix remodelling activities including the production of collagen. Furthermore, CAF MMP2 expression has a direct effect on TGF-β1 expression, which is implicated in both CAF and cancer cell behaviours facilitating metastasis [87].

## 3. Cancer Cell Interactions with the Immune System

Interactions between cancer cells and the immune system play an important role in both cancer progression and anti-cancer response. The myriad interactions between cancer and the immune system can be broadly grouped into four distinct categories: immunosurveillance, the anti-cancer immune response, immunosuppression, and cancer assistance. Through immunosurveillance, the healthy immune system constantly surveys tissues for signs of cancer, specifically the presence of tumour antigens such as oncoviral, mutated, or abnormally expressed proteins. Upon successful identification of tumour cells, the anti-cancer immune response proceeds as mediated by helper and killer T cells, which identify and lyse the cancer cells. However, in many cases, cancer cells and certain cells of the tumour microenvironment, such as CAFs, avoid or suppress this immune response, by preventing the proliferation of helper and killer T cells or by promoting the inflammation-mediated recruitment of immunosuppressive regulatory T cells (Treg) and myeloid-derived suppressor cells (MDSCs) [5,88]. Finally, in many cases, cancer cells are actually able to co-opt the activities of the immune system to aid tumour progression and even metastasis. Indeed, certain immune cells, such as macrophages, dendritic cells (DCs), lymphocytes, neutrophils and natural killer (NK) cells are prevalent at the metastatic site and are actively involved in promoting or suppressing cancer dissemination [89–91].

*Interplay between tumour, stroma, and immune system can affect tumour progression*

| Cell Type | Anti-Cancer Activities | Pro-Cancer Activities |
|---|---|---|
| 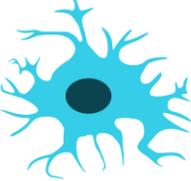 Dendritic Cell | - Adaptive immune response<br>- Present Tc cells with tumour antigens to target anti-tumour immune response<br>- Hypoxic conditions cause upregulation of T-cell costimulatory molecules and Th1/Th17-priming cytokines | - Heightened beta-catenin expression inhibits cross-priming of T cells with tumour antigens |
| 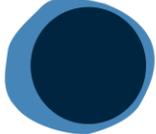 Cytotoxic T Cell (Tc CD8+) | - Cytotoxic immune response against cells expressing specific target antigens | - Cytotoxic immune response inhibited by cancer signalling<br>- Induced to apoptosis by cancer-initiated PD-1 signalling<br>- Produce cytokines including IL-17, promoting angiogenesis and tumour progression |
| 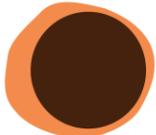 T Regulatory Cell (Treg, CD4+) | - Regulate adaptive immune response<br>- Control peripheral immune tolerance by suppressing Tc cell activity | - Cancer cell signalling causes suppression of the anti-tumor Tc response |
| 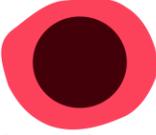 Myeloid-Derived Suppressor Cell (MDSC) | - Regulate immune response of cells including dendritic cells, T cells, and macrophages<br>- Possess a metastasis-suppressing phenotype | - Suppress the anti-tumour Tc cell immune response<br>- Potential to induce EMT<br>- Possess a metastasis-promoting phenotype |
| 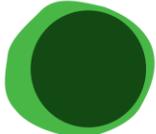 Natural Killer (NK) Cell | - Innate immune response<br>- Cytotoxic immune response against infection or tumor formation | - Cytotoxic immune response inhibited by cancer signalling |
| 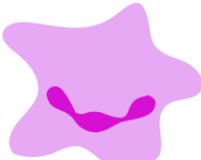 Macrophage | - Can adopt anti-inflammatory phenotype<br>- Innate immune response against infection<br>- Limited anti-tumour functionality | - Can adopt pro-inflammatory phenotype<br>- Secrete cytokines, survival, and growth factors<br>- Immunosuppressive signalling |
| 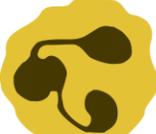 Neutrophil | - Can adopt anti-tumour phenotype<br>- Innate immune response against infection<br>- Maintain inflammatory response<br>- Secrete immunoactivating cytokines | - Can adopt pro-tumour phenotype<br>- Secrete cytokines, survival, and growth factors<br>- Immunosuppressive signaling<br>- Can inhibit NK cell cytotoxic function |

**Figure 3. Summary of selected immune cells' aberrant functions in cancer.** For each cell type, the associated immune subsystem (adaptive or innate immune response), aberrant function in cancer, and whether the disruption is from an increase or disruption in normal function is given.

*Interplay between tumour, stroma, and immune system can affect tumour progression*

*3.1. Macrophages*

Macrophages, which derive from monocytes after they enter the tissue from the circulation, constitute an important part of the body's primary immune response to infection and have also demonstrated anti-tumour functionality under normal circumstances (M2 phenotype) [92]. However, in many cancers, macrophages invade the tumour site and can adopt a tumour-promoting TAM phenotype (M1) in response to hypoxic, inflammatory, and otherwise abnormal tumour microenvironment. In a misplaced immune injury-healing response, TAM signalling promotes vascularization, invasiveness, growth, cancer cell survival, and immunosuppression, all resulting in continued tumour progression [92]. For instance, high levels of IL-10 expressed by TAMs in lung cancer are correlated with late stages of the disease, characterized by lymph node metastases, pleural invasion, lymphovascular invasion, and poor differentiation. These TAMs have a number of effects, from modulating inflammation and adaptive Th2 immunity to engaging in anti-inflammatory and tissue modelling activities [93]. This suggests that IL-10 is a powerful immunosuppressive factor that may promote cancer progression through its suppression of healthy macrophage immune activity [93]. TAMs (specifically the CD45+ M2 macrophages expressing fibroblast activation protein-$\alpha$ (FAP-$\alpha$)) have also been observed to maintain an immunosuppressive tumour microenvironment through their expression of heme oxygenase-1 (HO-1), an immune inhibitory enzyme which acts by suppressing the response of endothelial cells to immunogenic cytokines like tumour necrosis factor-$\alpha$ (TNF$\alpha$) [94].

*3.2. Neutrophils*

Constituting 50-70% of all circulating leukocytes, neutrophils are the most common type of white blood cell in circulation. They represent the traditional front line of defence against infection and are heavily involved in inflammation [95]. Throughout tumour progression, TANs transition from an anti-tumorigenic role characterized by the expression of immunoactivating cytokines and the production of TNF$\alpha$ toward tumour cell necrosis to a more pro-tumorigenic phenotype characterized by the expression of carcinogenesis, angiogenesis, and immune suppression factors (CXCR4, VEGF, MMP-9, arginase) [96]. These two subpopulations of anti-tumorigenic (N1) and pro-tumorigenic (N2) TANs both arise from the same initial population of normal neutrophils, suggesting that the changing tumour microenvironment is responsible for the difference in neutrophil phenotypes over time. In fact, when exposed to early-stage tumours, TANs largely confine themselves to the tumour periphery and adopt an anti-tumorigenic phenotype. However, when exposed to late-state, more developed tumours, neutrophils taken from the same initial population are found deeper in the tumour and furthermore adopt the aforementioned pro-tumorigenic phenotype [96]. This suggests that the degree of tumour development is the primary determinant of the resulting TAN phenotype. Additionally, it suggests that the ability to completely corrupt immune cells to the point of abandoning anti-tumour immune function is an evolved ability which is not immediately available to nascent tumours.

*3.3. T Cells*

T cells are a class of immune cells representing the key actors of the adaptive immune response. In the case of cancer, circulating tumour cell antigens are delivered to lymph nodes by DCs, where they are processed and presented to CD4+ and CD8+ T cells, activating these effectors commonly known as T helper (Th) and cytotoxic T killer (Tc) cells, respectively [97]. Upon leaving the lymph nodes, these circulating immune cells induce a cytotoxic immune response against cancer cells presenting their targeted antigens [97]. This process is further modulated by an immunosuppressive class of T cells known as T regulatory (Treg), which control peripheral immune tolerance, and thus possess immunosuppressive functionality. They are found in elevated numbers in the microenvironments of certain cancers, and play an important part in the suppression of the anti-tumour immune response [98].

*Interplay between tumour, stroma, and immune system can affect tumour progression*

In the course of disease progression, cancer cells develop a number of ways of avoiding and suppressing the T cell immune response. This includes the recruitment of immunosuppressive cells such as regulatory T cells and myeloid-derived suppressor cells, the evolution of the ability to induce apoptosis of activated T cells, and the development of various paracrine-signalling modalities which inhibit T cell activity [99,100]. For example, breast cancer cells presenting surface marker CD44 (CD44+ BCCs), have been shown to produce high levels of IL-8, granulocyte colony-stimulating factor (G-CSF), and TGF-β than cells not expressing CD44 (CD44- cells). Furthermore, these cells inhibit T cell proliferation, regulatory T cells, and myeloid-derived suppressor cells significantly more than CD44- cells. CD44+ BCCs are known to suppress the T helper (Th) cell immune response and enhance the immunosuppressive response of regulatory T (Treg) cells [101]. Lung cancer cells have been observed to express B7-homolog 4 (B7-H4), a homolog of B7.1/2 (CD80/86), which is known to have co-stimulatory and immune regulatory functions. The expression of this protein, induced by exposure to TAMs, on the cell membranes of lung carcinoma cells facilitated a powerful inhibition of T-cell action against the cells [102]. In hepatocellular carcinoma, T cells have been observed to express significantly more molecules of CD69, a leukocyte-expressed immunoregulatory molecule associated with chronic inflammation, than their non-tumour associated counterparts. Furthermore, in response to these CD69+ T cells, TAMs produced unusually high amounts of IDO protein, which in turn suppressed T cell immune responses in vitro [103].

As previously mentioned, when immune checkpoint receptor PD-1, expressed on T-cell membranes, is bound by its ligand PD-L1—as expressed on a cancer cell membrane—it causes T cell apoptosis [21]. However, when it occurs on DCs, this binding inactivates the T-cell immune response by inhibiting NF-kB activation in DCs, resulting in the suppression of NF-kB-dependent cytokine release and the inhibition of NF-kB mediated antigen presentation by the DCs [104]. A form of immunotherapy known as immune checkpoint blockade takes advantage of the frequency with which this immunosuppressive tactic is observed in cancer by blocking the signalling pathways associated with PD-1, PD-L1, and other immune checkpoint receptors [105]. By blocking cancer's ability to evade the immune response, these treatments allow the anti-cancer immune response to proceed [105].

More than simply avoiding the T cell response, cancer cells can also recruit T cells to act as tumour promoters. In response to IL-23, IL-6, and TGF-β in the tumour microenvironment, tumour-invading γδ T cells produce cytokine IL-17, which promotes angiogenesis and overall tumour promotion in animal models [106]. Furthermore, T-cells, in addition to NK cells, have been shown to promote metastasis to the lung [107].

*3.4. Dendritic Cells*

Cancer cells can also suppress and modulate the immune response indirectly. An important example of this is the cancer cell-mediated activation of β-catenin in DCs, the cells responsible for accumulating and presenting killer T cells with the specific tumour antigens responsible for directing the anti-tumour immune response. In response to higher levels of active β-catenin, DCs demonstrate an inhibited cross priming of CD8+ T cells with tumour antigens, thus dampening the entire CD8+ T cell-mediated anti-tumour immune response [108]. Furthermore, in hypoxic conditions, DCs express triggering receptors usually found on myeloid cells such as TREM-1, an immunoreceptor which amplifies inflammation [109]. TREM-1 engagement results in an upregulation of T-cell costimulatory molecules and an increased production of Th1/Th17-priming cytokines, all of which result in increased T-cell responses [109].

DCs also impact metastatic dissemination. In fact, lung resident DCs play a significant role in inhibiting metastasis of B16 melanoma [110]. Patrolling monocytes also participate in cancer surveillance by preventing tumour metastasis to the lung in a CX3CR1-dependent manner. These monocytes engulf tumour cells to prevent them from metastasizing, and also recruit and activate NK cells whose role is to secrete cytotoxic chemokines to destroy cancer cells [111].

*3.5. Myeloid-Derived Suppressor Cells*

Myeloid derived suppressor cells (MDSCs), a subset of immune cell derived from bone marrow in response to aberrant myeloplasmosis from cancer-driven perturbations in STAT/IRF-8 signalling, exhibit strong immunosuppressive behaviours [112]. Through the action of the transcription factor C/EBPβ, tumour cells secrete the necessary cytokines--including GM-CSF, G-CSF, and IL-6—required to spur the rapid generation of myeloid-derived suppressor cells, which, through the deregulation of their immunoregulatory function, are able to suppress the anti-cancer CD8+ T cell immune response [113]. Furthermore, TNF signalling through the TNF receptor-2 (TNFR-2) and the NF-kB pathway promotes the accumulation and survival of MDSCs by upregulating c-FLIP and inhibiting caspase-8 activity [114].

An interesting non-immunosuppressive function observed in MDSCs in the tumour microenvironment is their ability to induce the epithelial-mesenchymal transition in melanoma. Specifically, MDSCs infiltrate the primary tumour in response to the expression of the chemokine CXCL5 in the tumour microenvironment. Upon invasion, they induce EMT through the activation of the TGF-β, EGF, and HGF signalling pathways in cancer cells [115]. Importantly, this pro-metastatic function of MDSCs suggests a very direct link between inflammation, cancer-associated immune cells, and pro-metastatic tumour development. MDSCs have also been reported to adopt a metastatic-promoting or a metastatic-suppressing phenotype depending on their interactions with cancer-induced B regulatory cells (tBregs) [116].

**4. Interactions Between Immune Cells and the Tumour Stroma**

*4.1. Immune Cell Recruitment to Primary Tumours*

Since 1909 when Ehrlich predicted that immune cells act as a suppressor of cancer spreading, substantial research has emerged to question, re-evaluate or approve this theory [117]. In the past century, *in vivo* and *in vitro* work has shown that several types of cancer tissues, including breast cancer [118], glioma [119], and colorectal cancer tissues [120] contain higher fractions of specific immune cells than the normal tissue counterparts, suggesting that the immune system interacts very closely with the tumour microenvironment.

*4.1.2. Mechanochemical Properties of the Tumour ECM and Immune Cell Recruitment.* Rapidly growing tumours are known to generate hypoxia in their milieu which in turn promotes the angiogenesis of leaky vessels vascularizing the tumour microenvironment [121]. Additionally, ECM production from tumour growth generates mechanical stress on the leaky vessels which leads to an increase in interstitial fluid pressure and lymph flow. This flow delivers antigen-presenting cells (APCs) carrying antigens captured in the periphery of the tumour, and soluble pathogens and cytokines which promote the recruitment of macrophage and lymph node-resident DCs to the primary tumour site [43].

Amplified TGF-β production has been shown to attract NK cells, neutrophils and macrophages to the tumour microenvironment in preparation for an anti-tumour immune response [43]. However, the same TGF-β produced by the fibroblasts also neutralizes the anti-tumour functions of the recruited immune cells, thus favouring tumour evasion from the immune system [43]. In addition, TGF-β has been shown to block cytotoxic T-lymphocytes whose goal is to eradicate tumour cells, thus promoting immune evasion [122]. Recruited macrophages, activated by CAF-secreted CCL2 and upregulated SMAD protein signalling in tumour cells, often secrete additional TGF-β, further stimulating the production of collagen and collagen crosslinking enzyme LOX by the tumour stroma, which in turn promotes the fibrosis and stiffening of the ECM [33].

Tumour stroma stiffening has also been linked to an increase in the number of infiltrated TAMs in breast cancer [33]. In fact, CAFs and myofibroblasts associated with stiffened tumour ECM secrete soluble factors, such as chemokine ligand 2 (CCL2) and IL-1β, which stimulate macrophage recruitment [123]. These recruited macrophages are in turn stimulated by CCL2 and IL-1β to secrete the chemokine CXCL12 which induces angiogenesis to support the expanding tumour tissues [124]. From a mechanical point of view, the highly stiff tumour ECM affects the macrophages' adhesive interactions and cytoskeletal organization, enhancing their actin cytoskeletal contractility and promoting the development of a more elongated morphology and higher F-actin levels [125]. This may strengthens the rigidity of the cancer ECM even further, bolstering the malignant profile of the tumour.

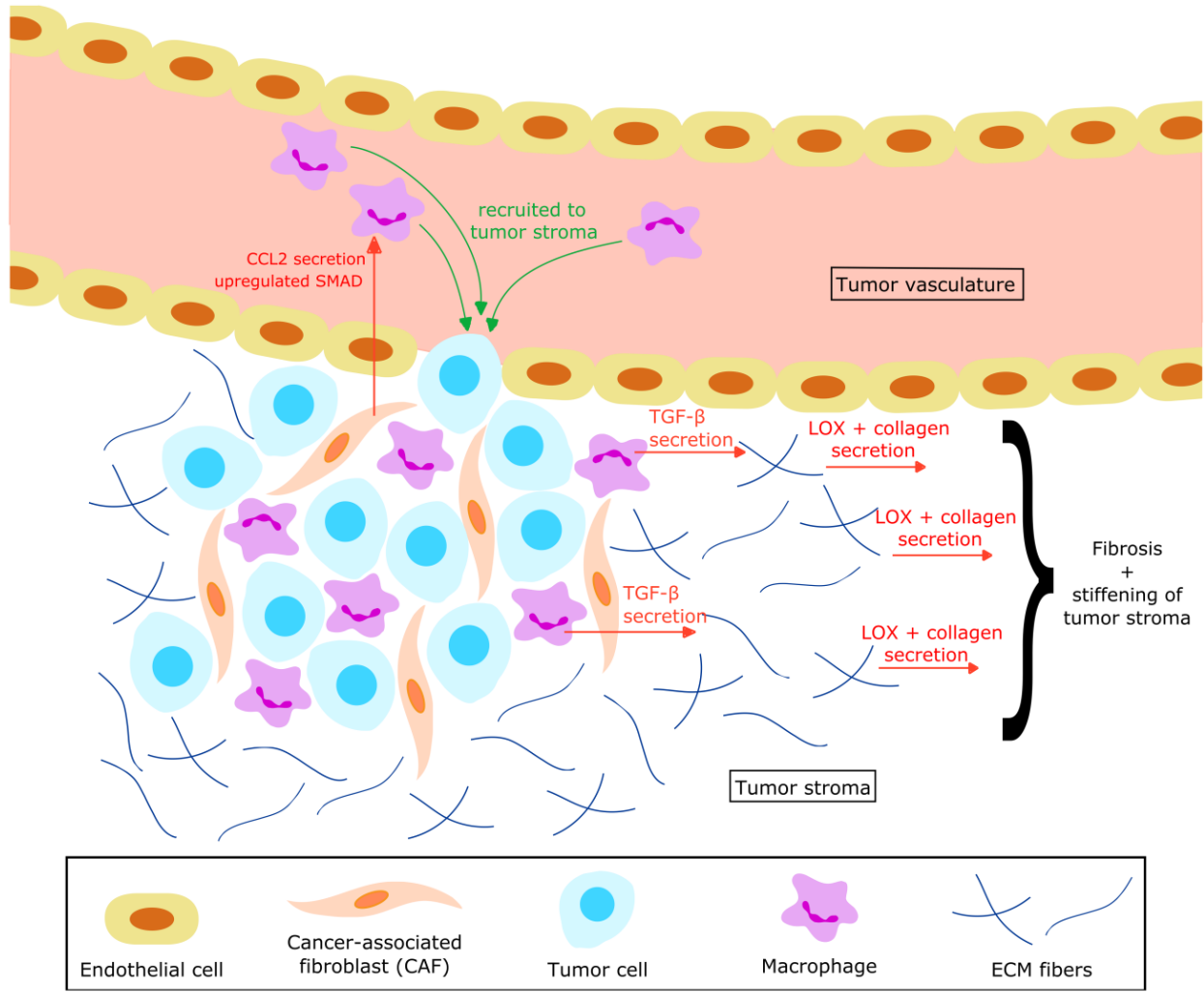

**Figure 4. Macrophage positive-feedback effect on tumour ECM stiffening.** Macrophages are recruited to the tumour stroma through CCL2 secretion and SMAD upregulation in CAFs. Once recruited, these macrophages secrete TGF-β which stimulates the production of collagen and LOX by the tumour stroma which in turn leads to fibrosis and tumour ECM stiffening.

*4.1.3. Oxygenation of the Tumour Stroma and Immune Cell Recruitment.* Hypoxia in the tumour microenvironment is responsible for the upregulation of heterodimer complexes called hypoxia-inducible factors (HIFs) which play a critical role in tumour immune responses. HIF1α, for example, has been shown to enhance the recruitment of macrophages and monocytes from the bone-marrow to the primary tumour site [126]. Previous reports have also suggested that hypoxia may favour the ability of DCs to provide signals to T-cells, alerting them to the presence of cancer cells and modulating their immune response by upregulating the expression of CD69 and CD141 receptors on the T-cells, thus priming them into pro-inflammatory T-cells [127]. Conversely, hypoxia has been reported to down-regulate anti-tumour T-cell responses in a different way, by favouring the accumulation of adenosine in the tumour microenvironment [126]. Interestingly, hypoxia acts in combination with specific immune cells in the tumour microenvironment to reduce immune response. For instance, reduced oxygen levels in the tumour stroma lead to increased inhibition of T-cell proliferation by macrophages, thus reducing their anti-tumorigenic response and enhancing tumour growth [128].

*4.1.4. Acidity of the Tumour Stroma and Immune Cell Recruitment.* Cancer cells perform fermentative glycolysis in their hypoxic milieus to generate adequate energy to sustain their growth. In terms of the immune response, this translates to increased concentrations of lactic acid which have been reported to decrease the cytotoxicity of T-cells and NK-cells [129], and to modulate DC function [130], through reduced interferon-$\alpha$ (INF-$\alpha$) production. In fact, tumour-cell derived lactic acid suppresses cytotoxic T-cell proliferation, impairs T-cell cytokine production, and eventually kills up to 60% of the T-cells after 24 hours [131]. In the case of DCs, previous work has shown that lactic acid inhibits DC differentiation in tumours and alters the antigen phenotype of DCs generating special tumour-associated DCs. For instance, lactic acid inhibits the expression of antigen CD1a by DCs but promotes the expression of CD14, whereas a reversed phenomenon is observed in the absence of lactic acid [132].

*4.2. Mechanically-Activated Signalling Pathways Between Primary Tumours and Immune Cells*

In the upstream segment of the immune response, molecules and surface markers such as TGF-$\beta$, CCL2, CCL21 and major histocompatibility complex (MHC) Class 1 are known to play a key role in mechanically-induced immune cell recruitment and activation. As was mentioned earlier, increased interstitial flow imposes shear stress on the cells and on the ECM fibres to which the cancer cells attach. These may trigger TGF-$\beta$1 expression through $\alpha$1$\beta$1 integrin signalling since the $\beta$1 integrins have been associated with tensional force generation in fibroblasts [133]. TGF-$\beta$1 then drives $\alpha$-SMA expression by the fibroblasts which in turn differentiate into myofibroblasts shown to align the matrix fibres, leading to fibrosis in the tumour microenvironment [43]. In addition, antigen-specific T-cells, activated macrophages and mature DCs in the tumour microenvironments prominently express PD-1 in contrast with the corresponding immune cells in normal tissues and peripheral blood vessels [134,135]. TGF-$\beta$ has been shown to function synergistically with PD-1 expressed by corresponding immune cells to signal for their suppression, thus enhancing anti-tumour immunity [43]. The mechanical stress applied on the CAFs, from the increase in interstitial flow and from the stretch on the tumour matrix due to its expansion, activates the fibroblasts focal adhesion kinase (FAK) leading to the secretion of CCL2 which acts to recruit macrophages to the tumour site [136].

Fibroblastic reticular cells, a specific type of lymph node stromal cells of the paracortical reticular meshwork, face significant levels of shear stress from the increased fluid pressure in the lymphatic vessels which promotes their secretion of cytokine CCL21 [137]. The overexpression of CCL21 promotes the recruitment of naïve T-cells expressing the receptor of CCL21, chemokine receptor 7 (CCR7), into the tumour microenvironment. In addition, CCL21 secretion has been shown to alter patterns of antigen presentation and lymphatic homing of DCs expressing CCR7 [43].

Changes in the arrangement and function of the lymphatic system in cancer can have an effect on tumour progression and immune cell function. In addition to simply providing a conduit for the trafficking of lymphatic fluid, immune cells, various signalling species, and other chemical actors, the lymphatic system can also be co-opted into aiding cancer expression by the unique mechanical and biological characteristics of the tumor microenvironment. In tumours themselves, the increase of mechanical forces generated by tumour growth and the disruption of normal lymphangiogenesis result in a lack of proper lymphatic circulation and function [138]. However, at the tumour periphery, increased vascular endothelial growth factor (VEGF)-C signalling results in the hyperplasia of the lymphatic vessels at the tumour periphery, facilitating significantly increased lymphatic flow rates away from the tumour as well as increased lymph node metastasis [139]. These physical and functional changes in the tumour-proximal lymphatic system, in addition to increasing the metastatic potential of the cancer cells, can also have an effect on immune activity. Under tumour conditions, PD-L1 signalling by lymphatic endothelial cells (LECs) can promote CD8+ T-cell tolerance, thus disrupting the anti-cancer immune response [140].

From a mechanical standpoint, heightened transmural lymph flow in the tumour microenvironment acts as a directional cue for DC trafficking [141]. On the one hand, increased flow promotes CCL21 secretion, a DC chemoattractant, as well as the secretion of DC adhesion molecules ICAM-1 and E-selectin, facilitating DC migration and crawling on the inner lymph vessels. On the other hand, heightened transmural lymph flow decreases lymphatic junctional adhesion molecules PECAM-1 and VE-cadherin, thus stimulating the lymphatic endothelium to increase fluid transport [142]. Furthermore, lymphatic endothelial cells (LECs) are able to actively seek out exogenous antigens within circulating lymph in preparation for presentation on MHC class I molecules, a class of cell surface markers which have been shown to play a role in tumour immune escape [143]. These molecules are involved in antigen presentation to cytotoxic T-cells and in the regulation of NK cell function [144]. In the tumour microenvironment, it has been shown that MHC class I molecules are downregulated on the surface of cancer cells, which allows T-cells and NK cells to recognize and eliminate these tumour cells ("missing self" hypothesis) [145,146]. In addition, increased solid stress and interstitial fluid pressure in the tumour microenvironment has been shown to induce mechanical stress on the cancer cells by physically shedding MHC class I on the cells surface [42,145]. This MHC class I downregulation in turn induces a positive cytotoxic response by NK cells [147].

*4.3. Dual Role of Immune Cells as Cancer Promoters or Cancer Suppressors*

*4.3.1. Immune Cells Known to Act as Promoters, Suppressors, or Both in the Tumour Microenvironment.* Immune cells possess a dual role in the tumour microenvironment. While certain tumour stroma characteristics can activate immune cells to become tumour suppressors, other tumour properties and signals can promote tumour evasion by the immune system [148–151]. In one example of this functional dichotomy, macrophages can polarize towards a pro-inflammatory (pro-tumour) M1 state in the presence of macrophage colony stimulating factor, TNF or IFN-γ produced at sites of inflammation, such as the tumour site, or, in response to IL-4, IL-10 or IL-13 they can polarize towards an M2, anti-inflammatory (anti-tumour) phenotype [148,152]. In addition, while most T-cells produce IFN-γ which plays an inhibitive role on cancer cells, specific subsets of T-cells, such as regulatory T-cells expressing the surface marker CD25 or Th17 producing IL-17, promote tumour growth by inhibiting the anti-cancer immune response [148]. TANs possess a polarized profile influenced by TGF-β in the cancer microenvironment. When recruited to the tumour site, TANs acquire an N2, pro-tumour phenotype enhanced by the presence of cancer cell-secreted TGF-β, but in its absence, they acquire an N1, anti-tumour state [151,153]. Neutrophils recruited to the pre-metastatic brain niche by the upregulation of S100A8 and S100A9 proteins secrete S100A4, which has been shown to promote glioma progression with mesenchymal characteristics, suggesting that neutrophils recruited to the brain tumour microenvironment acquire a pro-tumorigenic phenotype [154,155]. Finally, in the brain tumour microenvironment, glioma cells release colony-stimulating factor 1 (CSF-1), a chemoattractant for microglial cells which also converts their phenotype into a pro-tumorigenic one. Glioma cells also release CCL2 which acts on the CCL2 receptor (CCR2) expressed on microglia, triggering the release of IL-6 from microglia which makes glioma cells even more invasive [119,156].

*4.3.2. Mechanical Properties and Dual Role of Immune Cells.* The tumour microenvironment is characterized by increased blood flow and interstitial fluid pressure, which leads to elongated and dilated venules in the tumour vicinity as well as amplified shear stress on the vessel walls [157]. TGF-β, secreted from stiffened CAFs in the tumour stroma [43], in combination with IL-6, secreted in the tumour ECM in response to the continuous stress exerted on blood vessel endothelial cells [158], promotes the differentiation of T-cells into cancer-promoting Th17 cells [150,159]. The secretion of TGF-β1 by the fibroblasts through interstitial flow-induced α1β1 integrin signalling is also responsible for the differentiation of neutrophils into N2, pro-tumour neutrophils [43,153]. These responses favours immune evasion and tumour proliferation.

*4.3.3. Influence of Specific Molecules and Signalling Pathways of the Tumour Stroma to the Immune Cell Response.* At the molecular level, several receptors and molecules have been identified to play a key role in the dual role of immune cells as cancer promoters and cancer suppressors.

The secreted cytokines interferon-γ (IFN-γ) and toll-like receptor (TLR) agonists activate macrophages to adopt an M1 phenotype, thereby suppressing tumour growth [160]. However, in response to cancer cell-secreted IL-4 and IL-13, the same macrophages can promote tissue remodelling and cancer progression [161]. This macrophage phenotype depends on substrate stiffness and mechanically-activated signalling pathways resulting in the secretion of M1-stimulating molecules (such as IFN-γ and TLR) or M2-stimulating molecules (IL-4 and IL-13, among others), and the Rho-associated protein kinase II (ROCK II) is responsible for the switch in phenotype between M1 and M2 [123]. Increased substrate stiffness of the tumour microenvironment promotes the activation of ROCK II [162], which in turn increases the levels of IL-4, promoting an M2, pro-tumour phenotype [163].

Once macrophages are recruited to the tumour microenvironment, they also play a role in collagen remodelling at the tumour site, thus increasing the ECM stiffness even further. M2 pro-tumour macrophages induce the activity of the collagen cross-linker enzyme lysyl hydroxylase 2 (PLOD2) in the adjacent fibroblasts. TAMs also express the collagen cross-linker enzymes PLOD1 and PLOD3 which induce the synthesis of collagen types I, VI and XIV thus promoting matrix stiffening at the tumour site [164]. These interactions between macrophages and tumour matrix stiffness comprise a positive feedback loop: ECM stiffness promotes macrophage recruitment to the cancer site, and this recruitment induces the production and deposition of collagen, further increasing the rigidity of the tumour ECM.

Neutrophil polarization is also dictated by a number of key molecular mechanisms which promote either an antitumor N1 or pro-tumour N2 phenotype. For example, stiffened CAFs have the tendency to secrete larger quantities of TGF-β, which is responsible for the recruitment and activation of N2 pro-tumour neutrophils [153], but also suppresses N1 antitumor neutrophil cytotoxicity [165].

*4.4. Immune cells and the metastatic cascade*

Recently, more emphasis has been placed on the role of immune cells in the metastatic cascade. While it has been reported that T-cells and NK cells can restrict the progression of breast cancer cells away from the primary tumour site [166], these same immune cells have also been shown to promote metastasis to the lung [107]. Macrophages have also been shown to act as both promoters and suppressors of metastatic dissemination. For instance, sub-capsular sinus (SCS) macrophages can create a barrier to block tumour-derived extracellular vesicles from migrating to form metastases [167]. Alternatively, macrophages have also been shown to foster angiogenesis and metastasis in the case of breast cancer [168]. Specifically, M2 macrophages play a significant role in promoting gastric and breast cancer metastases through the secretion of the CHI3L1 protein [169]. MDSCs have also been reported to adopt a metastatic-promoting or a metastatic-suppressing phenotype depending on their interactions with cancer-induced B regulatory cells (tBregs) [116].

Furthermore, neutrophils have been extensively studied in the context of secondary tumours. They have been shown to facilitate the metastatic dissemination as well as the intermediate steps of the metastasis cascade of murine mammary carcinoma 4T1 cells. Neutrophils produce these effects by inhibiting NK cell cytotoxic function, which increases the survival time of intraluminal MDA-MB-231 breast cancer cells in 4T1 mice, and by promoting extravasation through the secretion of MMPs and IL-1$\beta$, which facilitate cancer cell dissemination and activates endothelial cells to secrete specific MMPs, respectively [148]. Finally, neutrophils act as mediators in the "two-step adhesion" process through which they bind to the endothelium and also to melanoma cells, thus facilitating the extravasation of tumour cells by promoting their adhesion to the ECM. Both endothelial cells and melanoma cells express ICAM-1 on their surface which binds to $\beta 2$ integrins of neutrophils, which act as the missing link to facilitate cancer cell-tumour ECM adhesions [170].

*4.4.1. Mechanical Properties of the Metastatic Site and Immune Cell Behaviour.* The mechanical properties of the ECM have a significant effect on cancer dissemination and immune cell-cancer cell interactions, and furthermore, immune cells have been shown to play a role in various steps of the metastatic cascade, from cancer cell detachment and dissemination from the primary tumour, to their arrest, extravasation and colonization of the metastatic sites [7,148,171]. Therefore, in the context of the tumour mechanical properties, the 'countercurrent' model has been developed to explain the mechanics of metastasis formation favoured by immune cell-cancer cell interactions [170]. For example, secreted chemokines from cancer cells and stromal cells of the primary tumour microenvironment favours neutrophil recruitment. These neutrophils play a role in promoting the loss of adhesion of cancer cells to their ECM, through a process called 'tumour shedding' which involves neutrophil-secreted molecules found in large quantities in neutrophil-dense tumour microenvironments. In fact, neutrophil-mediated extravasation occurs first by a tethering of neutrophils on the endothelium, followed by a capture of cancer cells to the neutrophils and maintenance of these cells near the endothelium to facilitate their extravasation [170].

Neutrophils also facilitate extravasation by degrading the matrix through their secretion of specific MMPs, thus creating a channel towards the cancer cell [148]. This channel may also be used by cancer cells that have lost adhesion contacts with their ECM to migrate in the opposite direction, towards the blood vessels from which the neutrophils are recruited. This 'countercurrent' model describes one way in which immune cell recruitment to the primary tumour site mechanically remodels the ECM to create a channel through which cancer cells can migrate to form metastases [170].

*4.4.2. Immune Cell Promotion of Metastases.* While neutrophils play a fundamental role in the various steps of the metastatic cascade, macrophages, lymphocytes, bone-marrow derived DCs and platelets have been shown to facilitate cancer cell dissemination from the primary tumour site and to protect circulating tumour cells from NK cell-mediated lysis and destruction from applied shear forces during their transit through the vasculature to the metastatic site [90].

In melanoma, SCS macrophages are recruited to the primary tumour site to bind to tumour-derived extracellular vesicles, creating a barrier around the tumour mass. However, with the applied interstitial pressure in the tumour vicinity, cancer cells proliferate and lymph nodes become enlarged without expanding their SCS macrophage pool. Unlike TAMs, SCS macrophages do not divide in the tumour microenvironment, resulting in a disrupted SCS barrier around the expanding tumour vesicles. Mechanically, this disrupted barrier creates breaches for the cancer cells from the primary tumour to escape and disseminate to secondary sites forming metastases [167].

Once cancer cells have travelled away from the primary tumour site, their arrest depends on the quality and quantity of their adhesion molecules and those found on the endothelial cells lining the vessel walls, as well as size restriction from narrow endothelial vessels [90]. Platelets and neutrophils have been shown to play a significant role in facilitating the extravasation of cancer cells by remodelling the matrix and its attachment molecules at the site of arrest [90]. In fact, in melanoma, platelets and neutrophils promote the production of matrix attachment molecules, such as $\beta 2$ integrin, intercellular adhesion molecule 1 (ICAM-1) and E-selectin, to facilitate cancer cell migratory arrest [172].

## 5. Outlook and Concluding Remarks

Cancer cells directly interact with the immune system through paracrine signalling and direct cell-cell contact signalling in order to suppress their anti-cancer function and promote tumour spreading and metastasis. Cancer cells also interact with their stroma, remodelling their ECM to facilitate their own invasion and metastatic dissemination. The cycle completes through interactions between the tumour stroma and the immune system wherein the stiff, hypoxic and inflamed tumour microenvironment affects immune cell behaviour. Taken as a whole, this tripartite interaction between cancer cells, immune cells and tumour stroma contributes to the maintenance of a chronically inflamed tumour microenvironment with pro-tumorigenic immune phenotype and facilitated metastatic dissemination.

Understanding these interactions requires ongoing experimental work and mathematical modelling—a particularly useful tool in understanding such complex systems. Models of immune system-cancer interactions have incorporated a wide range of mathematical approaches, including differential equations, cellular automata, and spatial and non-spatial multiscale models [173–175]. These models have examined tumour progression in the immune context [176–178], cancer cell-immune cell interactions in general [179,180], and immune suppression and escape [181]. Furthermore, some have incorporated microenvironmental or stromal elements such as paracrine signalling [182], stiffness sensing [183], or considered cancer-immune interactions as they effect therapy response [184,185]. Though the tri-directional interactions between cancer, the immune system, and the tumour stroma have been explored in some models, mathematical modelling efforts will continue to build an updated understanding of this dynamic interplay.

With a better understanding of the microenvironmental factors affecting tumour progression, new opportunities for cancer therapies become apparent. Paracrine signalling is fundamental to cancer cells' influence on cancer-associated stromal cells and immune cells recruited to the tumour microenvironment, and by altering signalling pathways known to promote specific immune cell polarizations, their pro-tumorigenic function could be suppressed in the tumour stroma. Alternatively, treatments which simultaneously tackle the three components of the cancer tripartite interaction could more effectively kill tumour cells, or at least limit tumour progression and metastatic dissemination. Finally, by interrupting the modalities by which cancer remodels its microenvironment, much of the pro-tumour positive feedback described above may be prevented.

In summary, in this perspective we have endeavoured to demonstrate that the evolution of a tumour is dictated by more than the activities of cancer cells alone and is in large part determined by interactions with the immune system, many of which are mediated by their mutually surrounding stroma. Therefore, a comprehensive understanding of cancer in the context of the dynamic interplay of immune system and the tumour stroma is required to truly understand the progression toward and past malignancy, and furthermore, such an understanding is essential to developing effective cancer therapies which treat the entirety of the tumour phenotype.

**Acknowledgements**

The authors declare no conflicts of interest.

RJS, FS, and MHZ acknowledge support from the National Cancer Institute (U01CA177799, 1P01HL120839, and 1U01 CA202123-01).

CH and RDK acknowledge support from the National Cancer Institute (CA202177).